\documentclass{ieeemag}

\usepackage{amsmath}
\usepackage{amssymb}
\usepackage{amsthm}
\usepackage[usenames]{pstcol}
\usepackage{pstricks,pst-node,pst-tree}
\usepackage{multirow}
\usepackage[usenames]{color}

\title{Spam: It's Not Just for Inboxes and Search Engines!\\Making Hirsch h-index Robust to Scientospam}

\theoremstyle{definition}

\input{psfig.sty}
\input{srhindex.mac}

\renewcommand{\baselinestretch}{1}

\begin{document}

%%%%%%%%%%%-------------authors
\author{\sf Dimitrios Katsaros$^{1,2}$  \ \ \ Leonidas Akritidis$^{1}$ \ \ \ Panayiotis Bozanis$^{1}$\\
        \small \sf $^{1}$Dept. of Computer \& Communication Engineering, University of Thessaly, Volos, Greece\\
        \small \sf $^{2}$Dept. of Informatics, Aristotle University, Thessaloniki, Greece\\
        \small \sf dimitris@delab.csd.auth.gr, \{leoakr,pbozanis\}@inf.uth.gr
        }

%%%%%%%%%%%------------date
\date{}

%%%%%%%%%%%------------maketitle
\maketitle

%%%%%%%%%%%------------abstract 
\begin{abstract}
What is the `level of excellence' of a scientist and the real impact of his/her work upon the scientific 
thinking and practising? How can we design a fair, an unbiased metric -- and most importantly -- a metric
robust to manipulation?
\end{abstract}

%%%%%%%%%%%-------------document text

\renewcommand{\baselinestretch}{1.05}
\normalsize
 
\section{Quantifying an individual's scientific merit}

%%%%%%%%%%%%%%%% Summary of what we are going to deal with
The evaluation of the scientific work of a scientist has long attracted significant
interest, due to the benefits by obtaining an unbiased and fair criterion. A few years
ago such metrics were yet another topic of investigation for the scientometric community
with only theoretical importance, without any practical extensions.

%%%%%%%%%%%%%%%% Is it important?
Very recently though the situation has dramatically changed; an increasing number of academic 
institutions are using such scientometric indicators to decide faculty promotions. Automated
methodologies have been developed to calculate such indicators~\cite{Ren-CACM07}. Also, funding
agencies use them to allocate funds, and recently some governments are considering the consistent 
use of such metrics for funding distribution. For instance, the Australian government has
established the {\it Research Quality Framework} (RQF) as an important feature in the fabric of
research in Australia\footnote{http://www.uts.edu.au/research/policies/resdata/RQF.html};
%~\cite{RQF-07};
the UK government has established the {\it Research Assessment Exercise} (RAE) to produce quality
profiles for each submission of research activity made by institution\footnote{http://www.rae.ac.uk}.
%~\cite{RAE-08}.

The use of such indicators to characterize a scientist's merit is controversial, and a plethora
of arguments can be stated against their use. In his recent article, David Parnas~\cite{Parnas-CACM07}
described the negative consequences to the scientific progress caused by the ``publish or perish"
marathon run by all scientists.

Following the reasoning of the phrase attributed to A.\ Einstein that ``Not everything that can be 
counted counts, and not everything that counts can be counted.", we stress that the assessment of a 
scientist is a complex social and scientific process that is difficult to narrow it into a single 
scientometric indicator. Most of the times, the verbal descriptions of a scholar's quality is probably 
the best indicator. Though, the expressive and descriptive power of numbers (i.e., scientometric 
indicators) can not unthinkingly be ignored; instead of devaluing them, we should strive to develop 
the ``correct set" of indicators and, most importantly, to use them in the right way.

%%%%%%%%%%%%%%%% Relevant work
No matter how skeptical is someone against the use of such indicators, the impact of a scholar can
quite safely be described in terms of the acceptance of his/her ideas by the wider scientific community
that s/he belongs to. Traditionally, this acceptance is measured by the number of authored papers and/or
the number of citations. The early metrics are based on some form of (arithmetics upon) the total number
of authored papers, the average number of authored papers per year, the total number of citations, the
average number of citations per paper, the mean number of citations per year, the median citations per
paper (per year) and so on. Due to the power-law distribution followed by these metrics, they present
one or more of the following drawbacks (see also~\cite{Hirsch-PNAS05}):
\begin{itemize}
\addtolength{\itemsep}{-0.5\baselineskip}
\item They do not measure the impact of papers.
\item They are affected by a small number of ``big hits" articles.
\item They have difficulty to set administrative parameters.
\end{itemize}

%%%%%%%%%%%%%%%% The introduction of the h-index
J.~E.~Hirsch attempted to collectively overcome all these disadvantages and proposed a pioneering 
metric, the now famous \hi~\cite{Hirsch-PNAS05}. \hi index was a really path-breaking idea, and
inspired several research efforts to cure various deficiencies of it, e.g., its aging-ignorant 
behaviour~\cite{Sidiropoulos-Scientometrics07}.

%%%%% What is the main problem?
Nevertheless, there is a latent weakness in all scientometric indicators developed so far, either
those for ranking individuals or those for ranking publication fora, and the \hi is yet another victim
of this complication. The inadequacy of the indicators stems from the existence of what we term here
--- for the first time in the literature --- the {\it scientospam}.

\section{The notion of scientospam}

With a retrospective look, we see that one of the main technical motivations for the introduction of
the \hi, was that the metrics used until then (i.e., total, average, max, min, median citation count)
were very vulnerable to self-citations, which in general are conceived as a form of ``manipulation".
In his original article, Hirsch made specific mention about the robustness of the \hi with 
respect to self-citations and indirectly argued that \hi can hardly be manipulated. Indeed 
\hi is more robust than traditional metrics, but it is not immune to them~\cite{Schreiber-epl07}.
Actually, none of the existing indicators is robust to self-citations. In general, the issue of
self-citations is examined in many studies, e.g., \cite{Hellsten-Scientometrics07}, and the usual
practise is to ignore them when performing scientometric evaluations, since in many cases it may
account for a significant part of a scientist's reputation~\cite{Fowler-Scientometrics07}.

At this point, we argue that there is nothing wrong with self-citations; they can effectively 
describe the ``authoritativeness" of an article, e.g., in the cases that the self-cited author 
is a pioneer in his/her field and s/he keeps steadily advancing his/her field in an step-by-step
publishing fashion, until gradually other scientists discover and follow his/her ideas.

In the sequel we will exhibit that the problem is much more complex and goes beyond self-citations;
it involves the ground meaning of a citation. Consider for instance the citing patterns appearing
in Figure~1.

\begin{figure}[!hbt]
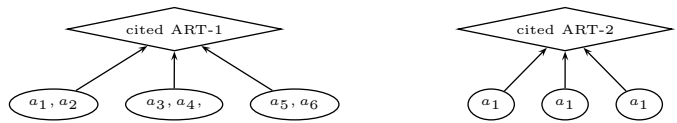

\centering
\begin{minipage}{2in}
\begin{center}
\psset{linewidth=0.5pt}
\psset{arrows=>}
\pstree[treesep=10pt,levelsep=1cm]{\Tdia{\tiny cited ART-1}}
       {
         \pstree{\Toval{\tiny $a_1,a_2$}} {}
         \pstree{\Toval{\tiny $a_3,a_4,$}} {}
         \pstree{\Toval{\tiny $a_5,a_6$}} {}
       }
\end{center}
\end{minipage}
\begin{minipage}{2in}
\begin{center}
\psset{linewidth=0.5pt}
\psset{arrows=>}
\pstree[treesep=10pt,levelsep=1cm]{\Tdia{\tiny cited ART-2}}
       {
         \pstree{\Toval{\tiny $a_1$}} {}
         \pstree{\Toval{\tiny $a_1$}} {}
         \pstree{\Toval{\tiny $a_1$}} {}
       }
\end{center}
\end{minipage}
\vspace*{-\baselineskip}
\caption{\sf \small Figure 1. Citing extremes: (Left) No overlap at all. (Right) Full overlap.}
\label{fig-citation-extremes}
\end{figure}

Article-1 is cited by three other papers (the ovals) and these citing articles have been authored
by (strictly) discrete sets of authors, i.e., $\{a_1,a_2\}$, $\{a_3,a_4\}$ and $\{a_5,a_6\}$, respectively.
On the other hand, Article-2 is cited by three other papers which all have been authored by the same 
author $\{a_1\}$. Notice that we make no specific mention about the identity of the authors of Article-1
or Article-2 with respect to the identity of the authors $a_i$; some of the authors of the citing papers
may coincide with those of the cited articles. Our problem treatment is more generic than self-citations.

While we have no problem to accept that Article-1 has received three citations, we feel that Article-2 has
received no more than one citation. Reasons to have this feeling include for instance the heavy influence
of Article-2 to author $a_1$ combined with the large productivity of this author. Nevertheless, considering
that all authors $a_1$ to $a_6$ have read (have they?) Article-1 and only one author has read Article-2, it seems
that the former article has a larger impact upon the scientific thinking. On the one hand, we could argue that
the contents of Article-2 are so sophisticated and advanced that only a few scholars, if any, could even grasp
some of the article's ideas. On the other hand, for how long could such situation persist? If Article-2
is a significant contribution, then it would get, after some time, its right position in the citation network,
even if the scientific subcommunity to which it belongs is substantially smaller that the subcommunity of
Article-1.

The situation is even more complicated if we consider the citation pattern appearing in Figure~2, where there
exist overlapping sets of authors in the citing papers. For instance, author $a_3$ is a coauthor in all three
citing papers.

\begin{figure}[!hbt]
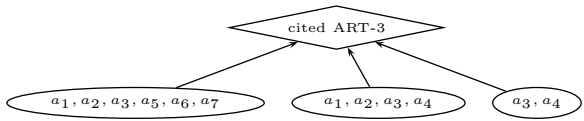

\begin{center}
\psset{linewidth=0.5pt}
\psset{arrows=>}
\pstree[treesep=10pt,levelsep=1cm]{\Tdia{\tiny cited ART-3}}
       {
         \pstree{\Toval{\tiny $a_1,a_2,a_3,a_5,a_6,a_7$}} {}
         \pstree{\Toval{\tiny $a_1,a_2,a_3,a_4$}} {}
         \pstree{\Toval{\tiny $a_3,a_4$}} {}
       }
\end{center}
\vspace*{-\baselineskip}
\caption{\sf \small Figure 2. Citing articles with author overlap.}
\label{fig-citation-overlap}
\end{figure}

This pattern of citation, where some author has coauthored multiple papers citing another paper
is the spirit of what is termed in this article the {\it scientometric spam} or {\it scientospam}.
The term spam is used in another two cases; it defines malicious emails ({\it e-mail spam}) and also 
Web links ({\it link spam}) that attempt to mislead the search engines when the engines exploit
some form of link analysis ranking. Whereas the word spam has received a negative reputation
representing malicious behaviour, we use it here as a means to describe misinformation.

Apparently, there exists no prior work on combating scientospam; the closest relevant works include
techniques to filter self-citations or weigh multi-author
self-citations~\cite{Schreiber-epl07,Schubert-Scientometrics06}.
Our target is to develop a metric of scientific excellence for individuals that will be really
robust to scientospam. We firmly believe that the exclusion of self-citations is not a fair action;
neither is any form of ad hoc normalization. Each and every citation has its value, the problem
is to quantify this value.

The notion of scientospam leads naturally to the process of the discovery of {\it spamming patterns} 
and their ``controlled discount". If we look more carefully at the citation data, we can gain a
deeper knowledge and thus produce a fairer and more robust evaluation. A more careful look implies 
that we have to pay some more computational cost than that for simple indicators, like \hi, but in
general we are willing to pay it, since the evaluation is an offline process. On the other hand, we
have to avoid time-consuming and doubtful clustering procedures and special treatment of self-citations,
so as to maintain the indicators' simplicity and beauty.

\section{The $f$-index}

We consider the citing example shown in Figure~2 where an article, say $\cA$, is cited by three other
articles and let us define the quantity $nca^{\cA}$ to be equal to the number of articles citing article \cA.
We define the series of sets  $F_i^{\cA}=\{a_j:\mbox{\it author $a_j$ appears in exactly $i$ articles citing \cA}\}$.
For the case of article ART-3, we have that $F_1^{\cA}=\{a_5,a_6,a_7\}$, $F_2^{\cA}=\{a_1,a_2,a_4\}$, $F_3^{\cA}=\{a_3\}$.

Then, we define $f_i^{\cA}$ to be equal to the ratio of the cardinality of $F_i^{\cA}$ to the total number of
distinct authors citing article \cA, i.e., $f_i^{\cA}= {\frac{|F_i^{\cA}|}{\mbox{\sf \tiny total \#distinct authors}}}$.
These quantities constitute the coordinates of a $nca^{\cA}$-dimensional vector $f^{\cA}$, which is equal to
$f^{\cA}=\{f_1^{\cA},f_2^{\cA},f_3^{\cA},\ldots,f_{nca^{\cA}}^{\cA}\}$. The coordinates of this vector define
a probability mass, since $sum_{i=1}^{nca^{\cA}} f_i^{\cA} =1$.
For the above example of the cited article ART-3, we have that $f^{ART-3}=\{\frac{3}{7},\frac{3}{7},\frac{1}{7}\}$
Similarly, for the cited article ART-1, we have that $f^{ART-1}=\{\frac{6}{6},\frac{0}{6},\frac{0}{6}\}$
and for ART-2, we have that $f^{ART-2}=\{\frac{0}{1},\frac{0}{1},\frac{1}{1}\}$.

Thus, we have converted a scalar quantity, i.e., the number of citations that an article has received, into a vector
quantity, i.e., $f^{\cA}$, which represents the penetration of \cA's ideas --- and consequently of its author(s) ---
to the scientific community; the more people know a scholar's work, the more significant s/he is. In general,
these vectors are sparse with a lot of $0$'s after the first coordinates. The sparsity of the vector reduces
for the cited articles which have only a few citations. Naturally, for successful scholars we would prefer the
probability mass to be concentrated to the first coordinates, which would mean that consistently new scientists
become aware of and use the article's ideas. As the probablity mass gets concentrated on the coordinates near
the end of $f^{\cA}$, the ``audience" gets narrower and it implies the existence of {\it cliques}, and/or 
citations due to {\it minimum publishable increment}, as they are both described by Parnas~\cite{Parnas-CACM07}.

Though, working with vectors is complicated and a single number would be the preferred choice. At this point,
we can exploit a ``spreading" vector, say $s$, to convert vector $f$ into a single number through a dot-product
operation, i.e., $\wf= f \cdot s$. For the moment will use the plainest vector defined as
$s_1=\{nca,nca-1,\ldots,1\}$; other choices will be presented in the sequel.
Thus, for the example article ART-3 which we are working with, we compute a new decimal number characterizing its
significance, and this number is equal to
$N_f^{\cA}= f^{\cA} \cdot s_1 = \frac{3}{7}*3 + \frac{3}{7}*2 + \frac{1}{7}*1 = \frac{16}{7} \Rightarrow N_f^{\cA} \approx 2.28$.

\subsubsection{The $f$-index.}
Now, we can define the proposed \fin in a spirit completely analogous to that of \hi. To compute the \fin of an
author, we calculate the quantities $N_f^{\cA_i}$ for each one of his/her authored articles $\cA_i$ and rank them
in a non-increasing order. The point where the rank becomes larger than the respective $N_f^{\cA_i}$ in the sorted
sequence, defines the value of \fin for that author.

\subsubsection{The spreading vector.}

Earlier, we used the most simple spreading vector; different such vectors can disclose
different facts about the importance of the cited article. Apart from $s_1$, we propose also a couple of easy-to-conceive
versions of the spreading vector. The vector $s_2= \{nca,0,\ldots,0\}$ lies at the other extreme of the spectrum
with respect to $s_1$. Finally, if we suppose that the last non-zero coordinate of $f^{\cA}$ is $f_k^{\cA}$, then we
have a third version of the spreading version defined as $s_3 = \{nca,nca-\frac{nca}{k},nca-\frac{2*nca}{k},\ldots,1\}$.
For each one of these spreading vectors, we define the respective \fin as $f_{s_1}$, $f_{s_2}$, and $f_{s_3}$.
None of these three versions of the spreading vector, and consequently of the respective indexes, can be considered 
superior to the other two. They present merits and deficiencies in difference cases. For instance, the $f_{s_1}$ index
does not make any difference for large \hi values; for scientists with \hi smaller than~15, the obtained $f_{s_1}$ index 
can be as much as 50\% of the respective \hi.

\section{Validation}

As we stressed right from the beginning of the article, when it comes to characterize the entire professional life of
a scholar with a single number, things get really complicated. The validation of the usefulness of the proposed indexes
is not an easy task, given our respect to the principle that ``not everything that can be counted counts". This article 
aims at introducing the notion of scientospam and proposing method to combat it. The comments made in this article should
not harm the reputation and will not reduce the contributions of any mentioned scientist. We selected as input data to 
apply our ideas a number of computer scientists with high \hi (http://www.cs.ucla.edu/$\sim$palberg/h-number.html), who 
are beyond any question top-quality researchers.

Since the data provided by the aforementioned URL are not up-to-date and also they are faulty, we cleansed them first,
we kept the scientists with \hi larger than~30. The ranking in non-increasing \hi is illustrated in Table~1.

\begin{figure}[!hbt]
\begin{center}
\tiny
\begin{tabular}{||@{}l@{}||r@{--}c||@{}l@{}||r@{--}c||@{}l@{}||r@{--}c|}\hline
{\bf r}& {\bf Scientist}            &{\bf h}&{\bf r} & {\bf Scientist}          &{\bf h}& {\bf r} &          {\bf Scientist}            &{\bf h}\\\hline\hline
1  &      Hector Garcia-Molina 	&  77  	& 17 &         Oded Goldreich  	&  48 	& 23  &                    Carl Kesselman 	& 42 	\\\hline
2  &               Jiawei Han 	&  66  	& 17 &           Philip S. Yu  	&  48 	& 24  &                  Olivier Faugeras 	& 41 	\\\hline
3  &               Ian Foster 	&  65  	& 17 &     Prabhakar Raghavan  	&  48 	& 25  &                     Teuvo Kohonen 	& 40 	\\\hline
4  &            Robert Tarjan 	&  64  	& 17 &         Leslie Lamport  	&  48 	& 25  &                        Amit Sheth 	& 40 	\\\hline
5  &           Rakesh Agrawal 	&  62  	& 17 &     Douglas C. Schmidt  	&  48 	& 25  &                    Craig Chambers 	& 40 	\\\hline
6  &           Jennifer Widom 	&  60  	& 18 &      Michael I. Jordan  	&  47 	& 25  &               Demetri Terzopoulos 	& 40 	\\\hline
6  &            Scott Shenker 	&  60  	& 18 &        Donald E. Knuth  	&  47 	& 25  &                David A. Patterson 	& 40 	\\\hline
7  &        Jeffrey D. Ullman 	&  59  	& 18 &           Ronald Fagin  	&  47 	& 25  &                     Philip Wadler 	& 40 	\\\hline
8  &           Deborah Estrin 	&  58  	& 18 &           Micha Sharir  	&  47 	& 25  &                     Jose Meseguer 	& 40 	\\\hline
9  &             David Culler 	&  56  	& 19 &         H. V. Jagadish  	&  46 	& 25  &                    George Karypis 	& 40 	\\\hline
9  &              Amir Pnueli 	&  56  	& 19 &          Mihir Bellare  	&  46 	& 26  &                Geoffrey E. Hinton 	& 39 	\\\hline
10 &             Richard Karp 	&  55  	& 19 &           Pat Hanrahan  	&  46 	& 26  &                      Stefano Ceri 	& 39 	\\\hline
10 &          Serge Abiteboul 	&  55  	& 19 &     Garcia Luna Aceves  	&  46 	& 26  &                 Leonard Kleinrock 	& 39 	\\\hline
11 &          David J. DeWitt 	&  54  	& 20 &       Michael Franklin  	&  45 	& 26  &                    Saul Greenberg 	& 39 	\\\hline
11 &        David E. Goldberg 	&  54  	& 20 &          Alex Pentland  	&  45 	& 26  &                       Judea Pearl 	& 39 	\\\hline
12 &             Anil K. Jain 	&  53  	& 20 &           Martin Abadi  	&  45 	& 26  &                        David Dill 	& 39 	\\\hline
13 &        Hari Balakrishnan 	&  53  	& 20 &       Andrew Zisserman  	&  45 	& 27  &                       Vern Paxson 	& 38 	\\\hline
13 &            Randy H. Katz 	&  52  	& 20 &    Thomas A. Henzinger  	&  45 	& 27  &                 John A. Stankovic 	& 38 	\\\hline
14 &             Takeo Kanade 	&  52  	& 20 &            Vipin Kumar  	&  45 	& 27  &                Krithi Ramamritham 	& 38 	\\\hline
14 &           Rajeev Motwani 	&  51  	& 20 &            Nancy Lynch  	&  45 	& 27  &                   Ramesh Govindan 	& 38 	\\\hline
15 &              Don Towsley 	&  50  	& 21 &     Christos Faloutsos  	&  44 	& 27  &                     Jon Kleinberg 	& 38 	\\\hline
15 &    Chr. H. Papadimitriou 	&  50  	& 21 &        Thomas S. Huang  	&  44 	& 28  &       Al. Sangiovanni-Vincentelli 	& 37 	\\\hline
15 &          Sebastian Thrun 	&  50  	& 21 &            Sally Floyd  	&  44 	& 28  &                  Edmund M. Clarke 	& 37 	\\\hline
15 &            Jack Dongarra 	&  50  	& 21 &           Robin Milner  	&  44 	& 29  &              Herbert Edelsbrunner 	& 36 	\\\hline
15 &              Ken Kennedy 	&  50  	& 21 &                Won Kim  	&  44 	& 29  &                    Richard Lipton 	& 36 	\\\hline
16 &            Didier Dubois 	&  49  	& 22 &      M. Frans Kaashoek  	&  43 	& 29  &                  Ronald L. Rivest 	& 36 	\\\hline
16 &              Lixia Zhang 	&  49  	& 22 &                 Kai Li  	&  43 	& 29  &                  Willy Zwaenepoel 	& 36 	\\\hline
16 &         Michael J. Carey 	&  49  	& 22 &          Monica S. Lam  	&  43 	& 29  &                        Jason Cong 	& 36 	\\\hline
16 &      Michael Stonebraker 	&  49  	& 22 &         Sushil Jajodia  	&  43 	& 30  &                     Victor Basili 	& 35 	\\\hline
16 &           Moshe Y. Vardi 	&  49  	& 22 &            Rajeev Alur  	&  43 	& 30  &                       Mario Gerla 	& 35 	\\\hline
16 &         David S. Johnson 	&  49  	& 23 &     Raghu Ramakrishnan  	&  42 	& 30  &               Andrew S. Tanenbaum 	& 35 	\\\hline
16 &          Ben Shneiderman 	&  49  	& 23 &         Barbara Liskov  	&  42 	& 31  &                      Maja Mataric 	& 33 	\\\hline
16 &           W. Bruce Croft 	&  49  	& 23 &          Tomaso Poggio  	&  42 	& 32  &                     John McCarthy 	& 32 	\\\hline
17 &       Mihalis Yannakakis 	&  48  	& 23 &          Victor Lesser  	&  42 	& 32  &                    David Haussler 	& 32 	\\\hline
17 &              Miron Livny 	&  48  	& 23 &          Joseph Goguen  	&  42 	& 33  &                     Stanley Osher 	& 31 	\\\hline
17 &            Luca Cardelli 	&  48  	& 23 &             Henry Levy  	&  42 	& 33  &                         Tim Finin 	& 31 	\\\hline
\end{tabular}
\end{center}
\caption{\sf \small Table 1. Computer scintists' ranking based on \hi.}
\label{table-h-index}
\end{figure}

Then, we applied the new indicators $f_{s_2}$ and $f_{s_3}$ and the results appear in Table~2. Both indicators
cause changes in the ranking provided by the \hi. As expected, the values of the $f_{s_2}$ index are significantly 
different than the respective \hi values. It is important to note, that these differences (and their size) appear
in any position, independently of the value of the \hi. If these differences concerned only the scientists with
the largest \hi, then we could (safely) argue that for someone who has written a lot of papers and each paper has
received a large number of citations, then some overlap citations and some self-citations are unavoidable. This 
is not the case though, and it seems that there is a deeper, latent explanation.

\begin{figure}[!hbt]
\begin{center}
\tiny
\begin{tabular}{|@{}l@{}||r@{--}c@{--}c||@{}l@{}||r@{--}c@{--}c||@{}l@{}||r@{--}c@{--}c|}\hline
{\bf r}  & {\bf Scientist}             & $f_{s_2}$ & $f_{s_3}$ &{\bf r}& {\bf Scientist}          & $f_{s_2}$ & $f_{s_3}$ & {\bf r}    &          {\bf Scientist}            & $f_{s_2}$ & $f_{s_3}$ \\\hline\hline
1  &      Hector Garcia-Molina 	& 68 	&	74	& 17 &        Donald E. Knuth 	&  41 	&	45	& 21  &               Geoffrey E. Hinton 	& 37 	&	37	\\\hline
2  &                Jiawei Han 	& 57 	&	63	& 17 &           Philip S. Yu 	&  41 	&	46	& 22  &                    Teuvo Kohonen 	& 36 	&	39	\\\hline
2  &                Ian Foster 	& 57 	&	62	& 18 &            Miron Livny 	&  40 	&	45	& 22  &                 Andrew Zisserman 	& 36 	&	41	\\\hline
3  &             Robert Tarjan 	& 56 	&	61	& 18 &          Luca Cardelli 	&  40 	&	46	& 22  &                   Sushil Jajodia 	& 36 	&	41	\\\hline
4  &             Scott Shenker 	& 54 	&	59	& 18 &           Ronald Fagin 	&  40 	&	45	& 23  &                    Joseph Goguen 	& 35 	&	40	\\\hline
5  &            Jennifer Widom 	& 53 	&	58	& 18 &         H. V. Jagadish 	&  40 	&	44	& 23  &                      Rajeev Alur 	& 35 	&	41	\\\hline
5  &         Jeffrey D. Ullman 	& 53 	&	55	& 18 &          Didier Dubois 	&  40 	&	44	& 23  &                    Philip Wadler 	& 35 	&	38	\\\hline
6  &              David Culler 	& 52 	&	53	& 18 &          Alex Pentland 	&  40 	&	43	& 23  &                       Amit Sheth 	& 35 	&	39	\\\hline
7  &            Deborah Estrin 	& 51 	&	56	& 18 &        Thomas S. Huang 	&  40 	&	42	& 23  &                      Nancy Lynch 	& 35 	&	42	\\\hline
7  &            Rakesh Agrawal 	& 51 	&	60	& 18 &            Sally Floyd 	&  40 	&	43	& 23  &                Leonard Kleinrock 	& 35 	&	38	\\\hline
8  &         David E. Goldberg 	& 50 	&	52	& 18 &           Robin Milner 	&  40 	&	42	& 23  &                      Vern Paxson 	& 35 	&	37	\\\hline
9  &              Richard Karp 	& 49 	&	55	& 18 &      M. Frans Kaashoek 	&  40 	&	41	& 23  &                John A. Stankovic 	& 35 	&	37	\\\hline
10 &           David J. DeWitt 	& 48 	&	51	& 18 &         Carl Kesselman 	&  40 	&	42	& 24  &                   Saul Greenberg 	& 34 	&	37	\\\hline
10 &         Hari Balakrishnan 	& 48 	&	52	& 19 &         Moshe Y. Vardi 	&  39 	&	46	& 24  &                     Stefano Ceri 	& 34 	&	37	\\\hline
11 &              Anil K. Jain 	& 47 	&	50	& 19 &           Martin Abadi 	&  39 	&	43	& 24  &               Raghu Ramakrishnan 	& 34 	&	40	\\\hline
11 &               Amir Pnueli 	& 47 	&	52	& 19 &     Christos Faloutsos 	&  39 	&	43	& 24  &               Krithi Ramamritham 	& 34 	&	38	\\\hline
11 &              Takeo Kanade 	& 47 	&	50	& 19 &     Mihalis Yannakakis 	&  39 	&	46	& 24  &                    Jon Kleinberg 	& 34 	&	36	\\\hline
12 &             Randy H. Katz 	& 46 	&	51	& 19 &          Mihir Bellare 	&  39 	&	45	& 25  &                  Ramesh Govindan 	& 33 	&	36	\\\hline
12 &               Lixia Zhang 	& 46 	&	48	& 19 &         Oded Goldreich 	&  39 	&	45	& 25  &                 Edmund M. Clarke 	& 33 	&	34	\\\hline
13 &               Don Towsley 	& 45 	&	49	& 19 &     Garcia Luna Aceves 	&  39 	&	43	& 26  &                      Judea Pearl 	& 32 	&	36	\\\hline
13 &           Serge Abiteboul 	& 45 	&	52	& 19 &                 Kai Li 	&  39 	&	41	& 26  &                   Richard Lipton 	& 32 	&	35	\\\hline
13 &          David S. Johnson 	& 45 	&	48	& 19 &         Barbara Liskov 	&  39 	&	40	& 26  &                 Ronald L. Rivest 	& 32 	&	34	\\\hline
14 &               Ken Kennedy 	& 44 	&	49	& 19 &          Tomaso Poggio 	&  39 	&	41	& 26  &                    Victor Basili 	& 32 	&	35	\\\hline
14 &            Rajeev Motwani 	& 44 	&	48	& 19 &             Henry Levy 	&  39 	&	40	& 26  &              Andrew S. Tanenbaum 	& 32 	&	34	\\\hline
14 &           Sebastian Thrun 	& 44 	&	48	& 19 &       Michael Franklin 	&  39 	&	42	& 26  &                   David Haussler 	& 32 	&	34	\\\hline
14 &           Ben Shneiderman 	& 44 	&	48	& 20 &                Won Kim 	&  38 	&	42	& 27  &                    Jose Meseguer 	& 31 	&	37	\\\hline
14 &        Prabhakar Raghavan 	& 44 	&	46	& 20 &          Monica S. Lam 	&  38 	&	42	& 27  &                       David Dill 	& 31 	&	35	\\\hline
15 &            W. Bruce Croft 	& 43 	&	46	& 20 &            Vipin Kumar 	&  38 	&	41	& 27  &                 Willy Zwaenepoel 	& 31 	&	34	\\\hline
15 &     Chr. H. Papadimitriou 	& 43 	&	47	& 21 &          Victor Lesser 	&  37 	&	41	& 28  &      Al. Sangiovanni-Vincentelli 	& 30 	&	34	\\\hline
15 &         Michael I. Jordan 	& 43 	&	46	& 21 &    Thomas A. Henzinger 	&  37 	&	43	& 28  &                      Mario Gerla 	& 30 	&	33	\\\hline
16 &       Michael Stonebraker 	& 42 	&	45	& 21 &           Micha Sharir 	&  37 	&	43	& 29  &             Herbert Edelsbrunner 	& 29 	&	34	\\\hline
16 &             Jack Dongarra 	& 42 	&	48	& 21 &       Olivier Faugeras 	&  37 	&	40	& 29  &                        Tim Finin 	& 29 	&	30	\\\hline
16 &            Leslie Lamport 	& 42 	&	45	& 21 &         Craig Chambers 	&  37 	&	40	& 30  &                       Jason Cong 	& 28 	&	33	\\\hline
16 &        Douglas C. Schmidt 	& 42 	&	46	& 21 &    Demetri Terzopoulos 	&  37 	&	38	& 31  &                     Maja Mataric 	& 27 	&	30	\\\hline
16 &          Michael J. Carey 	& 42 	&	46	& 21 &     David A. Patterson 	&  37 	&	39	& 31  &                    Stanley Osher 	& 27 	&	31	\\\hline
16 &              Pat Hanrahan 	& 42 	&	44	& 21 &         George Karypis 	&  37 	&	38	& 32  &                    John McCarthy 	& 26 	&	29	\\\hline
\end{tabular}
\end{center}
\caption{\sf \small Table 2. Computer scintists' ranking based on $f_{s_2}$. The $f_{s_3}$ value is represented too.}
\label{table-f-indexes}
\end{figure}

Seeking this explanation, we calculated the differences in ranking positions for each scientist when ranked with \hi
versus when they are ranked with the $f_{s_2}$.\footnote{Scientists with the same \hi, have the same ranking position. 
For instance, J.\ Widom and S.\ Shenker each is ranked 6-th in the \hi ranking. The same holds for the ranking based 
on~$f_{s_2}$.} The results are illustrated in Table~3.

The general comment is that the scientists who climb up the largest number of positions are those whose work can ``penetrate"
(and thus benefit) large ``audiences". For instance, the research results by Lixia Zhang and John A. Stankovic, who work on 
sensors now, are cited in communities like databases, networking, communications. Other scientists whose works is used by
large audiences are those working on ``computer organization", e.g., M.\ Frans Kaashoek, Barbara Liskov, Andrew S. Tanenbaum, 
etc. Notice here, that scientists' age has nothing to do with the ranking relocation, since both younger researchers (e.g., 
Lixia Zhang) can climb up positions, just like elder scientists (e.g., Andrew S. Tanenbaum).

Another important question concerns whether the particular area of expertise of a researcher could help him/her 
acquire a
larger reputation. Undoubtedly, the research area plays its role, but it is not the definitive factor. Consider for
instance, the case of data mining which is a large area and has attracted an even larger number of researchers. We see
that George Karypis has earned four positions in the ranking provided by $f_{s_2}$. If the area of expertise was the
only rational explanation for that, then why Rakesh Agrawal, who founded the field, is among the scientists that lost 
the most number of positions in the ranking provided by $f_{s_2}$? The answers lies in the particularities of the 
research subfields; George Karypis contributed some very important results useful also in the field of bioinformatics. 
To strengthen this, we can mention the case of Jiawei Han. He is a data-mining expert whose work penetrates to 
communities like mining, databases, information retrieval, artificial intelligence, and his is ranked second, based
either on \hi, or on $f_{s_2}$ or on $f_{s_3}$.

\begin{figure}[!hbt]
\begin{minipage}{.5\textwidth}
\begin{center}
\tiny
\begin{tabular}{|@{}r@{--}c@{ }cc||rccc@{}|}\hline
{\bf Scientist}          & {\bf h}    & {\bf h-rank}  & {\bf earned pos. in $f_{s_2}$}\\\hline\hline
       David Haussler		&	32	&	32	&	+6	\\\hline
      Carl Kesselman		&	42	&	23	&	+5	\\\hline
    Geoffrey E. Hinton	&	39	&	26	&	+5	\\\hline
         Lixia Zhang		&	49	&	16	&	+4	\\\hline
     M. Frans Kaashoek	&	43	&	22	&	+4	\\\hline
      Barbara Liskov		&	42	&	23	&	+4	\\\hline
       Tomaso Poggio		&	42	&	23	&	+4	\\\hline
          Henry Levy		&	42	&	23	&	+4	\\\hline
      Craig Chambers		&	40	&	25	&	+4	\\\hline
    Demetri Terzopoulos	&	40	&	25	&	+4	\\\hline
    David A. Patterson	&	40	&	25	&	+4	\\\hline
      George Karypis		&	40	&	25	&	+4	\\\hline
         Vern Paxson		&	38	&	27	&	+4	\\\hline
     John A. Stankovic	&	38	&	27	&	+4	\\\hline
       Victor Basili		&	35	&	30	&	+4	\\\hline
   Andrew S. Tanenbaum	&	35	&	30	&	+4	\\\hline
           Tim Finin		&	31	&	33	&	+4	\\\hline
\end{tabular}
\end{center}
\end{minipage}
\begin{minipage}{.5\textwidth}
\begin{center}
\tiny
\begin{tabular}{|@{}r@{--}c@{ }cc||rccc@{}|}\hline
{\bf Scientist}    &{\bf h}	& {\bf h-rank}	& {\bf lost pos. in $f_{s_2}$}\\\hline\hline
Rakesh Agrawal		 &	62	&	5	  &	-2\\\hline
Amir Pnueli		     &	56	&	9	  &	-2\\\hline
Didier Dubois		   &	49	&	16	&	-2\\\hline
Mihalis Yannakakis &	48	&	17	&	-2\\\hline
Oded Goldreich		 &	48	&	17	&	-2\\\hline
Andrew Zisserman	 &	45	&	20	&	-2\\\hline
Jose Meseguer		   &	40	&	25	&	-2\\\hline
Serge Abiteboul		 &	55	&	10	&	-3\\\hline
Moshe Y. Vardi		 &	49	&	16	&	-3\\\hline
Micha Sharir		   &	47	&	18	&	-3\\\hline
Nancy Lynch		     &	45	&	20	&	-3\\\hline
\end{tabular}
\end{center}
\end{minipage}
\caption{\sf \small Table 3. Largest relocations w.r.t. rank position. (left) Most positions up. (Right) Most positions down.}
\label{table-largest-relocations}
\end{figure}

Examining the scholars with the largest loses, we see that scientists who have made ground-breaking contributions and
offered some unique results, e.g., Mihalis Yannakakis, and Moshe Y. Vardi, drop in the ranking provided by the $f_{s_2}$.
This has nothing to do with the theoretical vs.\ practical sides of the computer science; contrast the cases of 
M.\ Yannakakis and M.\ Vardi, versus A.\ Zisserman and R. Agrawal. It is due to the nature of the scientific results 
that do not ``resound" to other communities.

\section{Discussion}

When measuring science we should always have in mind the principle which says that ``not everything that can be counted
counts". On the other hand, we believe in the power of numbers and we side with Lord Kelvin which stated that
``When you can measure what you are speaking about, and express it in numbers, you know something about it. But when you
cannot measure it, when you cannot express it in numbers, your knowledge is of a meager and unsatisfactory kind: It may 
be the beginning of knowledge, but you have scarcely, in your thoughts, advanced to the stage of science."

We argue that instead of anathematizing each and every scientometric indicator, we should strive to develop the correct
set of them. David Parnas did an excellent job in recording a number of existing and significant problems with current
publication methodologies. Along the spirit of his ideas, we describe for the first time here, another dimension of
publication methodologies, the existence of {\it scientospam} and set forth an effort to discover the spamming patterns
in citation networks.

The astute reader will have realized by now that in our battle against the scientospam, we have in our arsenal the research
works dealing with Web link spam~\cite{Gyongyi-IEEEComputer05}, e.g., TrustRank, BadRank and so on. Unfortunately, the
situation is radically difficult in citation networks, because they consist of entities richer than the Web pages and the Web
links encountered in Web spam. Each node i.e., a citing article, in a citation network consists of entities i.e., co-authors,
which form a complex overlay network above the article citation network.

We believe that the detection of spamming patterns in citation networks is quite a difficult procedure, and the cooperation
of the authors is mandatory. Maybe the scientific community should set some rules about citing, rules not only ethical,
but practical as well. For instance, we could have sections in the ``References" section of each published article, to
describe which citations involve only relevant work, which citations refer to earlier work done by the authors of the 
article, which citations refer to works implemented as competing works in the article, and so on. Apart from these
organizational categories, others could be devised as well; whether the citing article's results contradict or support
the results of the cited articles and many other.

In any case, we believe that scientometric indicators are not a panacea, and we should work a lot before applying a set
of them to characterize the achievements of a scholar. Indicators do have their significance, but some methodologies, both
ethical and practical should change in order to have reliable and automated measurements of science.

%%%%%%%%%%%-------------bibliography
\bibliographystyle{plain}
\bibliography{srhindex}

\begin{thebibliography}{1}

\bibitem{Fowler-Scientometrics07}
J.~H. Fowler and D.~W. Aksnes.
\newblock Does self-citation pay?
\newblock {\em Scientometrics}, 72(3):427--437, 2007.

\bibitem{Gyongyi-IEEEComputer05}
Z.~Gy{\"o}ngyi and H.~Garcia-{M}olina.
\newblock Spam: {I}t's not just for inboxes anymore.
\newblock {\em IEEE Computer}, pages 28--34, October 2005.

\bibitem{Hellsten-Scientometrics07}
I.~Hellsten, R.~Lambiotte, and A.~Scharnhorst.
\newblock Self-citations, co-authorships and keywords: {A} new approach to
  scientists' field mobility.
\newblock {\em Scientometrics}, 72(3):469--486, 2007.

\bibitem{Hirsch-PNAS05}
J.~E. Hirsch.
\newblock An index to quantify an individual's scientific research output.
\newblock {\em Proceedings of the National Academy of Sciences},
  102(46):16569--16572, 2005.

\bibitem{Parnas-CACM07}
D.~T. Parnas.
\newblock Stop the numbers game: {C}ounting papers slows down the rate of
  scientific progress.
\newblock {\em Communications of the ACM}, 50(11):19--21, 2007.

\bibitem{Ren-CACM07}
J.~Ren and R.~N. Taylor.
\newblock Automatic and versatile publications ranking for research
  institutions and scholars.
\newblock {\em Communications of the ACM}, 50(6):81--85, 2007.

\bibitem{Schreiber-epl07}
M.~Schreiber.
\newblock Self-citation corrections for the {H}irsch index.
\newblock {\em Europhysics Letters}, 78(3), 2007.

\bibitem{Schubert-Scientometrics06}
A.~Schubert, W.~Gl{\" a}nzel, and B~Thijs.
\newblock The weight of author self-citations. {A} fractional approach fo
  self-citation counting.
\newblock {\em Scientometrics}, 67(3):503--514, 2006.

\bibitem{Sidiropoulos-Scientometrics07}
A.~Sidiropoulos, D.~Katsaros, and Y.~Manolopoulos.
\newblock Generalized {H}irsch h-index for disclosing latent facts in citation
  networks.
\newblock {\em Scientometrics}, 72(2):253--280, 2007.

\end{thebibliography}
%\nocite{*}

%%%%%%%%%%%-------------document end
\end{document}